\documentclass[prl,twocolumn,superscriptaddress]{revtex4}
\usepackage{amssymb} 
\usepackage{graphicx}
\usepackage{epsfig}  
\usepackage{subfigure} 
\usepackage{psfrag}

\begin{document} 
\begin{flushleft}
%SPhT-T05/127,   
%HU-EP-05/37,
%SFB/CPP-05-38
\end{flushleft}
 
\title{The Topology of Pseudoknotted Homopolymers} 
 
\author{Graziano Vernizzi} 
%\email{vernizzi@cea.fr} 
\affiliation{Service de Physique Th\'eorique, CEA Saclay, 91191 
Gif-sur-Yvette Cedex, France}  
 
\author{Paolo Ribeca} 
%\email{ribeca@physik.hu-berlin.de} 
\affiliation{Humboldt Universit\"at, Newtonstr.~15, 12489 Berlin, Germany}  

\author{Henri Orland} 
%\email{orland@cea.fr} 
\affiliation{Service de Physique Th\'eorique, CEA Saclay, 91191 
Gif-sur-Yvette Cedex, France}

\author{A. Zee} 
%\email{zee@itp.ucsb.edu} 
\affiliation{Department of Physics, University of California, Santa 
Barbara, CA 93106, USA} 
\affiliation{Institute for Theoretical Physics, University of 
California, Santa Barbara, CA 93106, USA} 
 
\begin{abstract} 
We consider the folding of a self-avoiding homopolymer on a lattice,
with saturating hydrogen bond interactions. Our goal is to numerically
evaluate the statistical distribution of the topological genus of
pseudoknotted configurations. The genus has been recently proposed for
classifying pseudoknots (and their topological complexity) in the
context of RNA folding.  We compare our results on the distribution of
the genus of pseudoknots, with the theoretical predictions of an
existing combinatorial model for an infinitely flexible and
stretchable homopolymer. We thus obtain that steric and geometric
constraints considerably limit the topological complexity of
pseudoknotted configurations, as it occurs for instance in real RNA
molecules. We also analyze the scaling properties at large homopolymer
length, and the genus distributions above and below the critical
temperature between the swollen phase and the compact-globule phase,
both in two and three dimensions.
\end{abstract} 
\maketitle 

%\section{Introduction}
One of the most exciting fields in modern computational molecular
biology is the search for tools predicting the complex foldings of
bio-polymers such as RNA \cite{SBS,tinoco,Higgs}, when homologous
sequences are not available.
\begin{figure}[t]
\centering 
\includegraphics[width=0.48\textwidth]{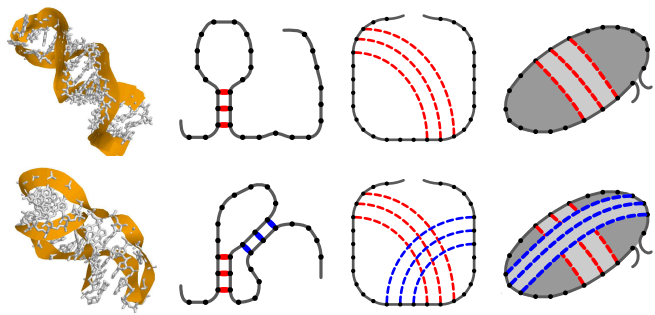}
\caption{Top, from left to right: 
%\cite{rasmol}: 
a hairpin loop (PDB number 1NA2), its squiggle-plot and disk diagram
representation which is of genus zero since it is planar.  Bottom:
$H$-type pseudoknot (PDB number 1RNK). In this case the disk diagram
is not planar and has genus one.}
\label{fig:1}
\end{figure}
The prediction of the full tertiary structure of a RNA
molecule is still an open issue \cite{mfold}, mainly because of its
intrinsic high computational complexity \cite{lyngso}. It is known 
that the tertiary structure involves an important set of
structural motifs, the so-called {\it pseudoknots}
\cite{Bosch}. These are conformations such that the
associated disk diagram  
(which represents all nucleotides along the RNA backbone as points on an
oriented circle from the 5' end to the 3' end, and where each base-pair is
represented by an arc joining the two interacting
nucleotides, inside the circle; see Figure \ref{fig:1}) is not planar, 
i.e.~it contains intersecting
arcs. RNA pseudoknots have been identified in nearly every organism,
and they proved to play important regulatory and functional roles 
\cite{c1,c2}. Their ubiquity manifests in a large variety
of possible shapes and structures \cite{plos}, and their existence
should not be neglected in structure prediction algorithms, as they
account for the 10\%-30\% on average of the total number of base
pairs. Actually, several computer programs have been proposed for
predicting RNA secondary structures including pseudoknots
\cite{eddyrivas, POZ, PTOZ, HI, Gul, abrah, Tab} (the list is not 
exhaustive), but the complexity of the problem and the approximations
involved are usually such that the issue is far from being solved
\cite{eddyBIO}.\\
An analytic mathematical tool which can fully describe any RNA contact
structure including all possible pseudoknots, appeared first in
\cite{OZ}. There, all  RNA disk diagrams are  considered as Feynman diagrams
of a suitable field theory of $N \times N$ Hermitian matrices (a
combinatorial tool borrowed from quantum field theory). The latter is
known to organize all the diagrams according to an asymptotic
$1/N^2$ topological expansion at large-$N$ \cite{thooft}. This
provides in fact a rigorous way to classify non-planar diagrams, and
therefore it induces a natural topological classification of
pseudoknots \cite{PTOZ}.  Namely, to any given pseudoknotted
configuration (and more generally, to any contact structure of an
heteropolymer with binary saturating interactions), one can associate
an integer number $g$, the {\it genus}. It is defined as the
topological genus of the associated disk diagram, i.e.~by $\chi=1-2g$
where $\chi$ is the Euler characteristic number of the diagram. As
reviewed in \cite{VOZ}, the genus is the minimum number of handles the
disk should have in order that all the cords are not intersecting (see
Figure \ref{fig:1}). Other characterizations of pseudoknots have been
proposed (e.g.~\cite{stella, Aeddy, Lucas}). The classification 
\cite{OZ} is truly topological, meaning that it is independent from the 
way the diagram is drawn, and dependent only
on the intrinsic complexity of the contact structure.

The large-$N$ asymptotics of the analytical model in \cite{OZ} is hard
to obtain exactly.  However, in \cite{PRL} a special case of the
general model \cite{OZ} has been considered and solved. It was the 
simple case of an infinitely flexible and stretchable homopolymer, where
there is no dependence on the primary sequence, and any saturating
base pair between all the ``nucleotides'' is allowed. An analytical
asymptotic expansion was evaluated and the distribution of the genus of
pseudoknotted contact structures was obtained. One of the
results is that an homopolymer with $L$ nucleotides has an average 
genus close to the maximal one, that is $L/4$. Of course, real RNA 
molecules are not infinitely flexible and stretchable homopolymers. 
It is customary to assume that the bases $i$ and $j$ can interact only 
if they are sufficiently far apart along the chain 
(e.g.~$\left|i-j\right|\geq4$,
\cite{mfold}) because of bending rigidity.  Moreover helices have a long
persistence length ($\sim 200$ base pairs) and this 
ne\-ces\-sa\-ri\-ly con\-strains the al\-lowed pairings even more.  
We expect that including 
all steric and geometrical constraints should considerably decrease
the genus of allowed pseudoknots, compared to the purely combinatorial
case \cite{PRL} where the actual three-dimensional conformation was
neglected. The purpose of this Letter is to numerically analyze the 
effects of
steric and geometric constraints on the genus distribution of
pseudoknots topologies in homopolymers, in the same spirit of
\cite{PRL}.
\smallskip

%\section{The Model}
\underline{The Model:} 
we model the system by considering a polymer on a cubic
lattice, i.e.~a self-avoiding random walk with short-range attractive
interaction \cite{degennes}.  A self-avoiding walk (SAW) is a sequence
of neighboring lattice sites $i=0,1,\ldots,n$ with coordinates $\{
\mathbf{r}_i \}$, such that the same lattice-site cannot be visited
more than once. This is a standard approach
in polymer physics (and RNA: see 
e.g.~\cite{stella, Lucas}). The attractive interaction is usually 
used to describe bad solvent quality, but in our case we insist
more on the saturating nature of the hydrogen bond interactions. Such a 
requirement is crucial here, since the concept of the topological genus
for a contact structure can be defined unambiguously only when the
interactions are saturating. One of the most natural ways to model 
the interaction is by considering a ``spin'' model (see 
e.g.~\cite{GOO,FS, Tiana}).  Strictly speaking, our
model is a variation of the standard $\theta$-polymer model, and
similar interaction models for RNA on the lattice have been already
proposed (e.g.~\cite{orl}).  To each vertex $i$ we
associate a unit spin $\mathbf{s}_i$ which represents the
nucleotide direction with respect to the backbone. The only
allowed directions for $\mathbf{s}_i$ are the lattice ones. Moreover,
the spins cannot overlap with the backbone because of the excluded
volume between the nucleotides and the backbone.
%, i.e.~$\mathbf{s}_i
%\cdot (\mathbf{r}_{i \pm 1}-\mathbf{r}_{i})\neq a$ where $a$ is the 
%lattice spacing. 
The saturating nucleotide-nucleotide interaction
occurs when two spins $\mathbf{s}_i,\mathbf{s}_j$ on neighboring
sites, $\left| \mathbf{r}_i-\mathbf{r}_j \right|=1$, are pointing to
each other. The energy of a configuration $\{\mathbf{r}_i,\mathbf{s}_i \}$ is thus defined by the
Hamiltonian:
\begin{equation}
{\mathcal H}=-\epsilon \sum_{i<j} \, 
\delta(\mathbf{r}_i+ \mathbf{s}_i - \mathbf{r}_j) 
\delta(\mathbf{s}_i+ \mathbf{s}_j) 
\delta(\left| \mathbf{r}_i - \mathbf{r}_{j}\right|-1) \,  ,
\label{Hamiltonian}
\end{equation}
where $\epsilon\geq 0$ is an effective hydrogen-binding energy, the
same for all monomers of the chain. Let us note that since we are not
aimidgng to set up a realistic lattice model for RNA-folding, but rather
to understand steric effects on the genus distributions of a
homopolymer, we do not take into account stacking energies.

The basic features of our model are clear:
At high temperatures, we expect the system to be in
a swollen SAW state (entropy dominated coil state), 
whereas at lower temperatures 
we expect a kind of ``compact globule''-like phase
\cite{degennes}. The transition temperature $T_\theta$
defines the so-called $\theta$-point. However, details on the
thermodynamics, kinetics, phase diagram, etc. can be rather complex
\cite{orl,Lucas}. We limit ourselves here only to the analysis of the
genus distribution of pseudoknotted structures for comparing the
effects of stericity constraints versus the purely combinatorial model
of \cite{PRL}.  All other considerations are postponed elsewhere.
\smallskip

%\section{The Method}
\underline{The Method:}
The numerical sampling of the statistical distribution ${\mathcal Z}=
\sum_{\mathrm{SAW},\{\mathbf{s}_i\}} \exp({-\mathcal H/k_B T})$, where
$k_B$ is Boltzmann's constant, $T$ is the absolute temperature, and
the sum is restricted to SAWs and configurations of spins $\{
\mathbf{s}_i\}$ satisfying the aforementioned constraints, is
im\-ple\-men\-ted by using the Monte Carlo Growth Method. It was
originally proposed by T. Garel and H. Orland in \cite{GO} and has
been applied to several statistical systems since then (see references
in \cite{HO}).  It
consists in starting with an ensemble of chains at equilibrium and
then growing each chain by adding one monomer at a time with a
probability proportional to the Boltzmann factor for the energy of the
chain. At each step the ensemble remains at equilibrium (a detailed
description of the algorithm with applications can be found in
\cite{HO}). It belongs to the family of so-called
``population Monte Carlo algorithms'' \cite{Iba}, where, contrary to
the ``dynamical'' Markov Chain Monte Carlo methods, the population is
fully grown and evolved, non-dynamically. 
%The main advantage is that,
%in principle, there are no slowing-down effects close to critical
%points. 
At high temperatures we considered populations with a 
variable number of chains in the range 10000-40000, and with a 
typical  length of $L=500$ monomers (up to $L=1200$ in some cases). 
Accuracy and statistical averages were computed by
taking several independent populations (of the order of 40). 
At low temperatures we considered populations of up to 100000 chains. 
All the simulations have also been performed on a square 
lattice in two dimensions.
\smallskip

\begin{figure}[hbt]  
\centering \includegraphics[width=0.48\textwidth]{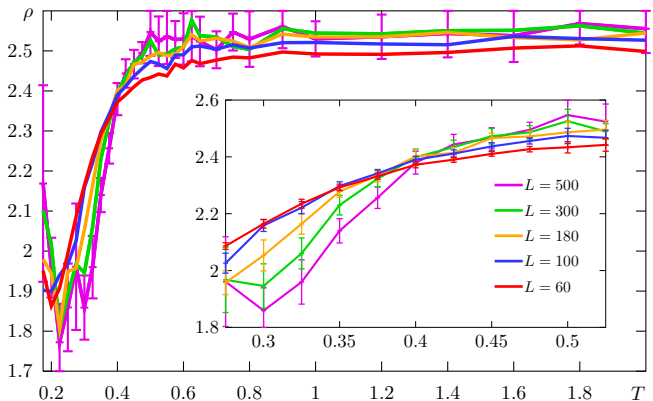}
\caption{The ratio $\rho$ in $3d$
as a function of the temperature $T$, for  several values
of $L$ (with error bars plotted only for $L=60$). At low temperatures
the error bars are larger than the variations of the curves.  
The inset shows that as $L$ increases,  the curves approach 
a  universal step function about $T_\theta\sim 0.39\pm 0.01$.}  
\label{fig:2}
\end{figure}

\begin{figure}[t]  
\centering 
\includegraphics[width=0.48\textwidth]{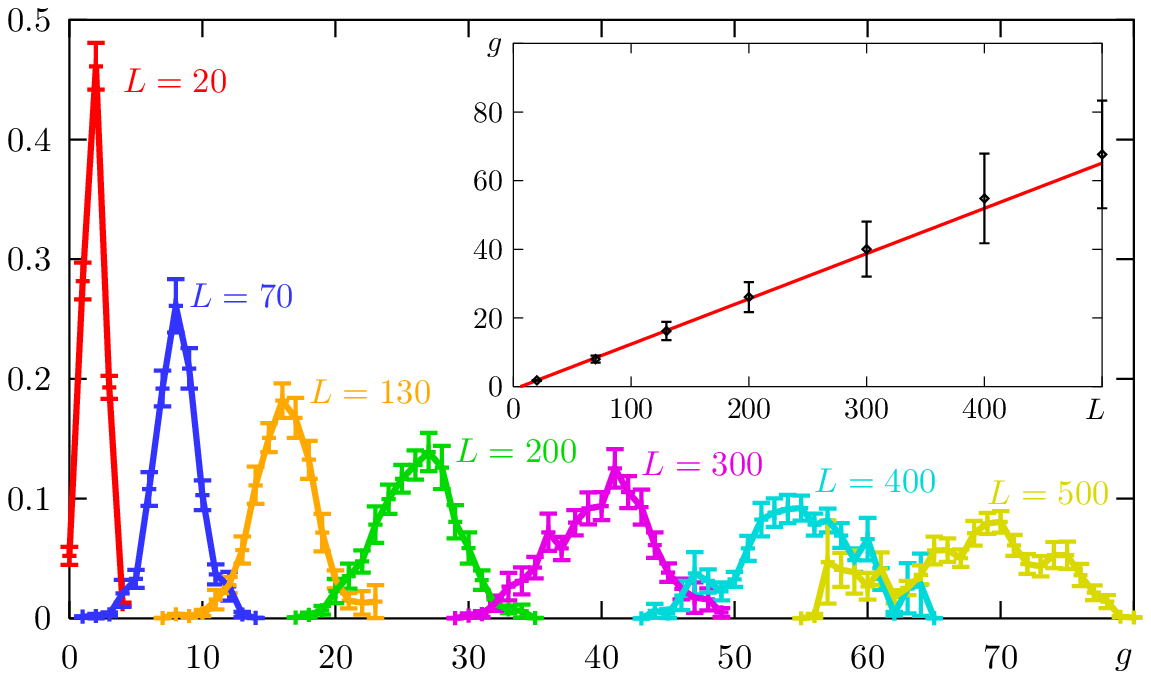}
\includegraphics[width=0.48\textwidth]{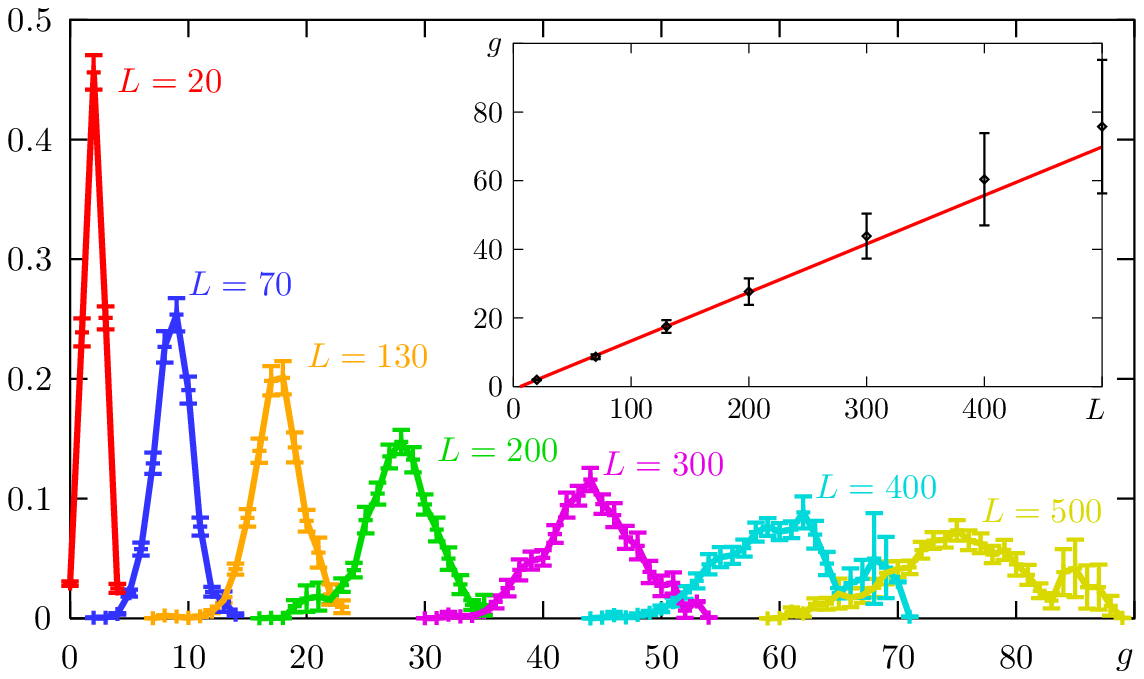}
\caption{The genus distributions  of pseudoknots at fixed 
$L$, in $2d$ (top) and $3d$ (bottom), 
in the compact phase at $T=0.225$ and $T=0.2$, respectively. 
The insets represent the 
behavior  of $\langle g\rangle$ at large-$L$.} 
\label{fig:3}
\end{figure}

\begin{figure}[hbt]
\centering
\includegraphics[width=0.48\textwidth]{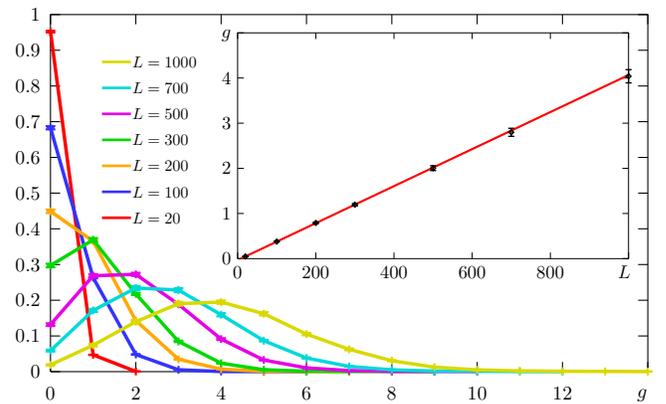}
\caption{The genus distribution of a two dimensional homopolymer, 
in the swollen phase $T=10>T^{2D}_\theta$, at various values of $L$.}
\label{fig:4}
\end{figure}

%\section{Results and discussion}
\underline{Results and discussion:}
We expect different genus distributions above and below $T_\theta$.
We therefore first determine $T_\theta$ , which can be done efficiently by
computing the end-to-end distance $R_e^2=(
 \mathbf{r}_L-\mathbf{r}_0)^2$, and the  radius of
gyration $R_g^2=\sum_{i<j} (\mathbf{r}_i-\mathbf{r}_j)^2/L^2$. 
It is known that the ratio $\rho^2=\langle R_e^2\rangle 
/\langle R_g^2 \rangle$ is universal in 
the limit $L \to \infty$ and converges to a step function as a
function of $T$, with a universal critical value at $T_\theta$
\cite{degennes, PHA, Nickel}. In Figure \ref{fig:2} we plot $\rho$ 
which shows a transition temperature $T_\theta=0.39 \pm 0.01$ ($
T^{2D}_\theta=0.48 \pm 0.02$ in two dimensions), in units 
where $k_B=1$ and $\epsilon=1$.
We also verified that $\rho_\infty=2.5\pm 0.05$ for $T\gg T_\theta$ 
asymptotically, 
and we find an intermediate value $\rho_\theta=2.35\pm 0.05$ at $T_\theta$ 
(and $\rho_\infty=2.67 \pm 0.01 $ and $\rho_\theta=2.39 \pm 0.01$ in 2d, 
respectively).  At 
large-$L$ we find the following scalings: for $T>T_\theta$, $\langle R_g 
\rangle  \sim L^\nu$ with $\nu=0.59 \pm 0.01$ ($\nu=0.75\pm 0.02$ in 
two dimensions), which is consistent with the critical exponent of a
swollen SAW; for $T<T_\theta$, $\nu=0.32 \pm 0.02$ ($\nu=0.50\pm 0.01$
in two dimensions) which is consistent with a compact phase.  All
these results are in agreement with high-accuracy simulations of
similar models \cite{grassberger,madras}.  We then proceed with
extracting the genus distributions in the two phases. The results are
in Figure \ref{fig:3} and \ref{fig:4}.
When comparing them with the combinatorial results of \cite{PRL}, we
see that the genus at a fixed $L$ is on the average much smaller.
More precisely, below the $\theta$-point the average genus scales like
by $\langle g / L \rangle \sim 0.141 \pm 0.003$ and $\langle g / L
\rangle \sim 0.1318\pm 0.0025$, in $3d$ (at $T=0.2$) and $2d$ (at
$T=0.225$), respectively. In both cases the scaling is at a lower rate
(about 50\% less) than the value $L/4$ computed in \cite{PRL}.  In the
swollen-phase (e.g.~$T=10>T_\theta$), the average genus is given by
$\langle g / L \rangle \sim (585 \pm 8)10^{-6}$ in $3d$, and $\langle
g / L \rangle \sim (410 \pm 1) 10^{-5}$ in $2d$.  Such a low rate
comes from the tendency of a homopolymer to develop long rectilinear
sub-chains in the swollen phase.  In two dimensions the entropic
factor is smaller than in three dimensions and the genus growth rate
is therefore larger (see Figure \ref{fig:4}).  Moreover, the genus
distributions for $T>T_\theta$ are numerically consistent with
Poissonian distributions (see Figure \ref{fig:4}), whereas at smaller
temperatures they are closer to Gaussian ones.

\begin{figure}[t]  
\centering
\includegraphics[width=0.48\textwidth]{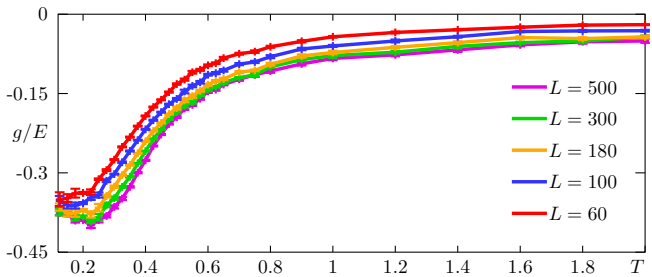}
 \caption{The ratio genus/energy of an homopolymer on a cubic lattice, as a 
function of $T$, at different values of $L$.}  
\label{fig:5}
\end{figure}

It turns out that the average genus is an extensive quantity, like the
energy, and their ratio is shown in Figure \ref{fig:5}.
All these results confirm that the genus distribution behaves
differently in the two phases, as expected.  They also quantify how
much the restrictions induced by the actual three-dimensional
arrangement of the chain can limit the number and complexity of
pseudoknots (compared to \cite{PRL}). We find values closer to what
seems to happen in pseudoknots of real RNA molecules. In fact, real
RNA molecules typically have small genus. For instance, a simple
$H$-type pseudoknot ($\sim$20 bases or more) or the classical
kissing-hairpins pseudoknot ($\sim$30 bases or more) both have genus
1, much less than the toy-model prediction in \cite{PRL}.  Even tRNAs
($\sim$80 bases) mostly contain 4 helices, two of them linked together
by a kissing-hairpin pseudoknot, has still genus 1. Typical tmRNAs
($\sim$350 bases long) contain four $H$-type pseudoknots, and its
total genus is 4, far below the theoretical upper bound $L/4$.  Our
numerical results would instead indicate, for instance for a $80$
bases long homopolymer, a genus of about $11.2$ in three dimensions
($\sim$10.5 in two dimensions). Even if it is smaller than the value
suggested in \cite{PRL} (because of the steric constraints), it is
still too high when compared to real RNA molecules.  The obvious
reasons are that we neither included the primary sequence nor
realistic stacking energies.  We have nevertheless been able to
quantify the general effect of steric constraints on the genus
distribution of a pseudoknotted homopolymer on a lattice, as a first
step towards a model which includes a more realistic energy
function.\\
\underline{Acknowledgements}: 
We wish to thank T. Garel and R. Guida  for  discussions.  
This work was supported in part by the National Science 
Foundation under grant number PHY 99-07949, and by 
Sonderforschungsbereich-Transregio ``Computational Particle Physics''
(SFB-TR9). GV acknowledges  the support of the European Fellowship 
MEIF-CT-2003-501547.

\end{document}